\documentclass{nature}

\usepackage{amsmath}
\usepackage{amssymb}
\usepackage{graphicx}
\usepackage{upgreek}

\title{A strong magnetic field around the supermassive black hole at the centre of the Galaxy}

\author{R.~P.~Eatough$^{1}$,
H.~Falcke$^{2,3,1}$, 
R.~Karuppusamy$^{1}$, 
K.~J.~Lee$^{1}$, 
D.~J.~Champion$^{1}$,
E.~F.~Keane$^{4}$,  
G.~Desvignes$^{1}$,
D.~H.~F.~M.~Schnitzeler$^{1}$, 
L.~G.~Spitler$^{1}$, 
M.~Kramer$^{1,4}$, 
B.~Klein$^{5,1}$, 
C.~Bassa$^{4}$,  
G.~C.~Bower$^{6}$,
A.~Brunthaler$^{1}$, 
I.~Cognard$^{7,8}$, 
A.~T.~Deller$^{3}$,
P.~B.~Demorest$^{9}$,
P.~C.~C.~Freire$^{1}$, 
A.~Kraus$^{1}$, 
A.~G.~Lyne$^{4}$, 
A.~Noutsos$^{1}$, 
B.~Stappers$^{4}$ \&  
N.~Wex$^{1}$ 
}

\begin{document}

\maketitle

\begin{affiliations}
 \item Max-Planck-Institut f\"ur Radioastronomie, Auf dem H\"ugel 69, D-53121 Bonn, Germany 
 \item Department of Astrophysics, Institute for Mathematics, Astrophysics and Particle Physics, Radboud University, PO Box 9010, 6500 GL Nijmegen, The Netherlands 
 \item ASTRON, P.O. Box 2, 7990 AA Dwingeloo, The Netherlands 
 \item Jodrell Bank Centre for Astrophysics, School of Physics and Astronomy, The University of Manchester, Manchester, M13 9PL, UK 
 \item University of Applied Sciences Bonn-Rhein-Sieg, Grantham-Allee 20, D-53757 Sankt Augustin, Germany 
 \item UC Berkeley Astronomy Dept, B-20 Hearst Field Annex, Berkeley, CA 94720-3411 
\item LPC2E / CNRS - Universit\'e d'Orl\'eans, 45071 Orl\'eans, France
\item Nan\c cay/Paris Observatory, 18330 Nan\c cay, France
\item National Radio Astronomy Observatory, 520 Edgemont Road, Charlottesville, VA 22903, USA 
\end{affiliations}

\begin{abstract}
The centre of our Milky Way harbours the closest candidate for a
supermassive black hole\cite{GenzelEisenhauerGillessen2010a}. The
source is thought to be powered by radiatively inefficient accretion
of gas from its environment\cite{NarayanYi1994}. This form of
accretion is a standard mode of energy supply for most galactic
nuclei. X-ray measurements have already resolved a tenuous hot gas
component from which it can be
fed\cite{BaganoffMaedaMorris2003a}. However, the magnetization of the gas, a crucial parameter
determining the structure of the accretion flow, remains
unknown. Strong magnetic fields can influence the dynamics of the
accretion, remove angular momentum from the infalling
gas\cite{BalbusHawley1991a}, expel matter through relativistic
jets\cite{BeckwithHawleyKrolik2008a} and lead to the observed
synchrotron
emission\cite{FalckeMarkoff2000,MoscibrodzkaGammieDolence2009,DexterAgolFragile2010a}.
Here we report multi-frequency measurements with several radio
telescopes of a newly discovered pulsar close to the Galactic
Centre\cite{2013arXiv1305.2128K,2013arXiv1305.1945M,eatough_atel1,sj+13}
and show that its unusually large Faraday rotation indicates a
dynamically relevant magnetic field near the black hole. If this field
is accreted down to the event horizon it provides enough magnetic flux
to explain the observed emission from the black hole, from radio to X-rays.
\end{abstract}

Linearly polarized radio waves that pass through a magnetized medium
experience Faraday rotation. The resulting rotation of the
polarization vector is given by $\Delta \phi = {\rm RM} \lambda^2$,
where the rotation measure ${\rm RM} = e^3/(2 \pi m_{\rm e}^2 c^4)\int
B(s) \, n(s) \,{\rm d}s$, which depends on the line-of-sight magnetic
field $B$, the free electron density $n$, and the path length $s$. The
radio emission associated with the Galactic Centre black hole,
Sagittarius~A* (Sgr~A*), has ${\rm RM}=-5\times10^5\;{\rm rad} \;{\rm
  m}^{-2}$, which is the highest known RM of any source in the Galaxy,
and is believed to be due to a column of hot, magnetized gas from the
accretion flow onto the black
hole\cite{BowerFalckeWright2005a,MarroneMoranZhao2007}.

Sgr A* itself, however, only probes the innermost scales of accretion.
For most models\cite{MarroneMoranZhao2007} the term $B(r)\,n(r)$ decays
faster than $r^{-1}$, where $r$ is the radial distance from the black
hole. Consequently, the Faraday rotation is dominated by the smallest
scales. To measure the magnetization of the accretion flow on the
outermost scales other polarized radio sources, such as pulsars, are
needed. A pulsar closely orbiting Sgr~A$^{\star}$ would also be an
unparalleled tool for testing the space-time structure around the
black hole\cite{lwk+12}. Despite predictions that there are in excess
of a thousand pulsars in the central parsec of the
Galaxy,\cite{wcc+12} there has been a surprising lack of
detections,\cite{ekk+13} potentially due to severe interstellar
dispersion and scattering in the inner Galaxy\cite{lc98}.

Recently the {\em Swift} telescope detected a bright X-ray
flare\cite{2013arXiv1305.2128K} near Sgr A$^{\star}$ ($\sim3''=0.12$
pc projected offset\cite{rea_atel} at a Galactic Centre distance $d=8.3$
kpc). Subsequent X-ray observations by the {\em NuSTAR} telescope
resulted in the detection of pulsations at a period of
$3.76$~s\cite{2013arXiv1305.1945M}. This behaviour is indicative of a
magnetar, a highly magnetized pulsar, in outburst. During radio
follow-up observations at the Effelsberg observatory on April 28th, a
first weak detection of pulsations, with spin parameters matching
those reported by {\em NuSTAR}, was made. Since then, the pulsar, PSR
J1745$-$2900, has been consistently detected at Effelsberg,
Nan\c{c}ay, the Karl G. Jansky Very Large Array (VLA), tentatively at
Jodrell Bank (see Figure 1.) and elsewhere with the
ATCA\cite{sj+13}. Measurements of the delay in the arrival times of
pulses at lower frequencies (2.5 GHz) with respect to higher frequency
(8.35 GHz) yield an integrated column density of free electrons, the
dispersion measure (DM), of $1778\pm3$ cm$^{-3}$ pc -- the highest
value measured for any known pulsar.  This is consistent with a source
located within $<10$ pc of the Galactic Centre, within the NE2001
density model of the Galaxy\cite{2002astro.ph..7156C}. Including this
source, only four radio-emitting magnetars are
known\cite{2012ApJ...744...97L} in the Milky Way, making a chance
alignment unlikely. If we consider a uniform source distribution
occupying a cylinder of radius 10 kpc and height 1 kpc, then the
fraction of sources present within an angular distance of $\sim3''$
around Sgr A$^{\star}$ is $\sim3\times10^{-9}$. Given the current
population of radio pulsars ($\sim$2000) and radio magnetars, the
number present within the same region by chance will be
$\sim6\times10^{-6}$ and $\sim1\times10^{-8}$ respectively.

The emission from the pulsar is highly linearly
polarized\cite{lee_atel,sj+13} (Figure 2.). Using the RM synthesis
method\cite{bb05} and measuring the Faraday rotation at three
frequency bands and three different telescope sites, we derive a RM of
$(-6.696 \pm 0.005) \times 10^{4} \; {\rm rad \; m^{-2}}$
(Figure. 3). This measurement is consistent with that presented
elsewhere\cite{sj+13}. The RM is the largest measured for any
Galactic object other then
Sgr~A$^{\star}$\cite{BowerFalckeWright2005a,MarroneMoranZhao2007}, and
is more than an order of magnitude larger than all the other RMs
measured to within tens of parsecs from
Sgr~A$^{\star}$\cite{LawBrentjensNovak2011a}. The RM is also more than
what can be optimistically expected as
foreground\cite{BowerBackerZhao1999a}. This constrains the Faraday
screen to be within some ten parsecs from the Galactic Centre.

A frequently used estimate of the magnetic field is ${\rm B} \ge {\rm
  RM}/(0.81 {\rm DM}) \upmu {\rm G}$, which
gives $B \ge 50 \, \upmu {\rm G}$\cite{sj+13}. However, this
is not a stringent limit, since the DM and RM are dominated by very
different scales. Hence, the extra information about the gas in the
central ten parsecs must be used for a more robust magnetic field
estimate.

There are two ionized gas phases in the Galactic Centre interstellar
medium towards the line of sight of the pulsar which could be
associated with the Faraday screen: a warm component from the Northern
Arm of the gas streamer Sgr A West\cite{ZhaoBlundellMoran2010a} that
passes behind Sgr~A$^{\star}$ and a diffuse hot component seen in the
X-rays\cite{BaganoffMaedaMorris2003a} with $T=2.2\times10^7$ K.  The
warm gas in the Northern Arm has a width of $>0.1$ pc, electron
densities of $\sim10^5\,{\rm cm}^{-3}$ measured from radio
recombination lines\cite{ZhaoBlundellMoran2010a}, and a magnetic field
of $\sim2$ mG\cite{PlanteLoCrutcher1995a}. The inferred RM and DM for
a source in or behind the Northern Arm are RM$\;\sim2\times 10^7 \;
{\rm rad \; m^{-2}}$ for an ordered magentic field and DM$\;\sim 10^4$ pc cm$^{-3}$. The measured DM and RM
values therefore place the pulsar and the screen {\it in
  front} of the Northern Arm\cite{ZhaoBlundellMoran2010a}.

Consequently, the Faraday screen must be associated with the hot gas
component, for which no magnetic field estimates exist yet. The
density in the hot gas shows a radial fall-off as a function of
distance $r$ from the black hole. At 0.4 pc ($10''$) one finds
$n\sim26\;{\rm cm}^{-3}$, whereas at 0.06 pc ($1.5 ''$) one can infer
$n \lesssim 160\;{\rm cm}^{-3}$, using the optically thin thermal
plasma model\cite{BaganoffMaedaMorris2003a}. Further away at the 40 pc
scale\cite{MunoBaganoffBautz2004a} ($17'$), the density has decreased
to 0.1--0.5 cm$^{-3}$ and we can roughly describe the central parsecs
with a density profile of the form $n(r)\simeq26\,{\rm
  cm}^{-3}\,(r/0.4\,{\rm pc})^{- 1}$.  The contribution of this hot gas
  component to the DM is of order $10^2$ cm$^{-3}$ pc.  This is
  consistent with the modest increase of DM with respect to the
  hitherto closest pulsars to the Galactic Centre.

  For a simple one-zone Faraday screen, where ${\rm RM} \propto B(r)
  \, n(r) \, r$, we have ${\rm RM}=8.1\times 10^5\,(B(r)/{\rm G})
  (n(r)/{\rm cm}^{-3}) (r/{\rm pc})\;{\rm rad} \;{\rm m}^{-2}$. Using
  the density prescription above with a $r^{-1}$ scaling, we find $B
  \gtrsim 8 ({\rm RM}/66960\,{\rm m^{-2}}) (n_0/26\;{\rm
    cm}^{-3})^{-1}$ mG. This is a lower limit, since possible
  turbulent field components or field reversals reduce the RM. We note
  again that this RM is indeed dominated by the smallest distance
  scale, i.e., by the gas on scales of the de-projected distance,
  $r>0.12$ pc, of the pulsar from Sgr A*.

  This value is higher than the magnetic field in the Northern Arm and
  also higher than the equipartition field in the hot phase at this
  scale. To bring thermal and magnetic energy into equipartition, the
  gas density at $r\sim0.12$ pc would need to increase by a factor of
  three to 260 cm$^{-3}$, yielding $\sim 2.6$ mG. Many field reversals
  within the Faraday screen would drive the magnetic field way beyond
  equipartition, suggesting that a relatively ordered magnetic field
  is pervading the hot gas close to the supermassive black hole.

  As Sgr~A$^{\star}$ accretes from this magnetized hot phase, density
  and magnetic field will further increase inwards. Emission models of
  Sgr A* require about 30-100 G magnetic fields to explain the
  synchrotron radiation from near the event
  horizon\cite{FalckeMarkoff2000,MoscibrodzkaGammieDolence2009,DexterAgolFragile2010a}.
  Hence, if the gas falls from $3\times10^5$ Schwarzschild radii (0.12
  pc) down to a few Schwarzschild radii, already a simple $B\propto
  r^{-1}$ scaling would be enough to provide several hundred Gauss
  magnetic fields. This is well within the range of most accretion
  models, where where equipartition between magnetic, kinetic, and
  gravitational energy in the accreting gas is
  assumed\cite{MacquartBowerWright2006,MarroneMoranZhao2007}.

The field provided on the outer scale of the accretion flow onto
Sgr~A$^{\star}$ is therefore sufficient to provide the necessary field on the
small scales via simple accretion. Moreover, the availability of
ordered magnetic fields would make the proposed formation of a
jet-like outflow in Sgr A*\cite{FalckeMannheimBiermann1993}
viable. Super-equipartition magnetic fields could also suppress
accretion and help to explain the low accretion rate of
Sgr~A$^{\star}$.

At its projected separation PSR J1745$-$2900 could move (due to
orbital motion) through the hot gas surrounding Sgr A* with several
mas$/$yr and reveal RM variations as well as proper motion. Continued
pulsar polarimetry and VLBI astrometry can readily measure these
effects. Also, given that magnetars constitute only a small fraction
of the pulsar population and the excess DM towards the Galactic Centre
is not too large, we expect additional observable radio pulsars to be
lurking in the same region.  Such pulsars could be used to map out the
accretion region around the black hole in more detail, and even test
its spacetime properties.










\begin{addendum}
 \item The authors wish to thank D.~D.~Xu., P.~Lazarus and
   L.~Guillemot for useful discussions. We also thank O.~Wucknitz and
   R.~Beck for manuscript reading. R.K., L.G.S. and
   P.C.C.F. gratefully acknowledge the financial support by the
   European Research Council for the ERC Starting Grant BEACON under
   contract no. 279702. K.J.L. was funded by ERC Advanced Grant LEAP
   under contract no. 227947. H.F. acknowledges funding from an
   Advanced Grant of the European Research Council under the European
   Union’s Seventh Framework Programme (FP/2007-2013) / ERC Grant
   Agreement no. 227610. This work was based on observations with the
   100-m telescope of the MPIfR (Max-Planck-Institut f\"ur
   Radioastronomie) at Effelsberg. The Nançay radio telescope is part
   of the Paris Observatory, associated with the Centre National de la
   Recherche Scientifique (CNRS), and partially supported by the
   R\'egion Centre in France. The National Radio Astronomy Observatory
   (NRAO) is a facility of the National Science Foundation operated
   under cooperative agreement by Associated Universities, Inc.

 \item[Correspondence] Correspondence and requests for materials
should be addressed to Ralph Eatough~(email: reatough@mpifr-bonn.mpg.de).\\
\end{addendum}

\noindent Author contributions:\\
R. P. Eatough - Initial detections, observations performed with Effelsberg, and data processing.\\
H. Falcke - Observational and theoretical background, paper formulation.  \\
R. Karuppusamy - Observational technical assistance and pulsar timing. \\
K. J. Lee - Polarization and RM measurements. \\
D. J. Champion - Pulsar timing solution. \\
E. F. Keane - Flux density calculations, observational assistance and observations at Jodrell Bank. \\
G. Desvignes - Observations Nan\c{c}ay. \\
D. H. F. M. Schnitzeler - Observational background, RM interpretation. \\
L. G. Spitler - Observational background, data processing and analysis. \\
M. Kramer - Observational background, RM interpretation. \\
B. Klein - technical observational assistance Effelsberg. \\
C. Bassa - Observations Jodrell Bank. \\
G. C. Bower - Observations VLA, RM interpretation. \\ 
A. Brunthaler - Observations VLA. \\
I. Cognard - Observations Nan\c{c}ay \\                        
A. T. Deller - Observations VLA \\                
P. B. Demorest - Observations VLA \\                
P. C. C. Freire - Observational background, pulsar timing. \\
A. Kraus - technical observational assistance Effelsberg. \\
A. G. Lyne - Observations Jodrell Bank, help with initial detections. \\
A. Noutsos - Observational background, RM interpretation.  \\
B. Stappers - Observations Jodrell Bank. \\
N. Wex - Theoretical background, orbital characteristics. \\

\begin{figure}
\includegraphics[scale=0.68]{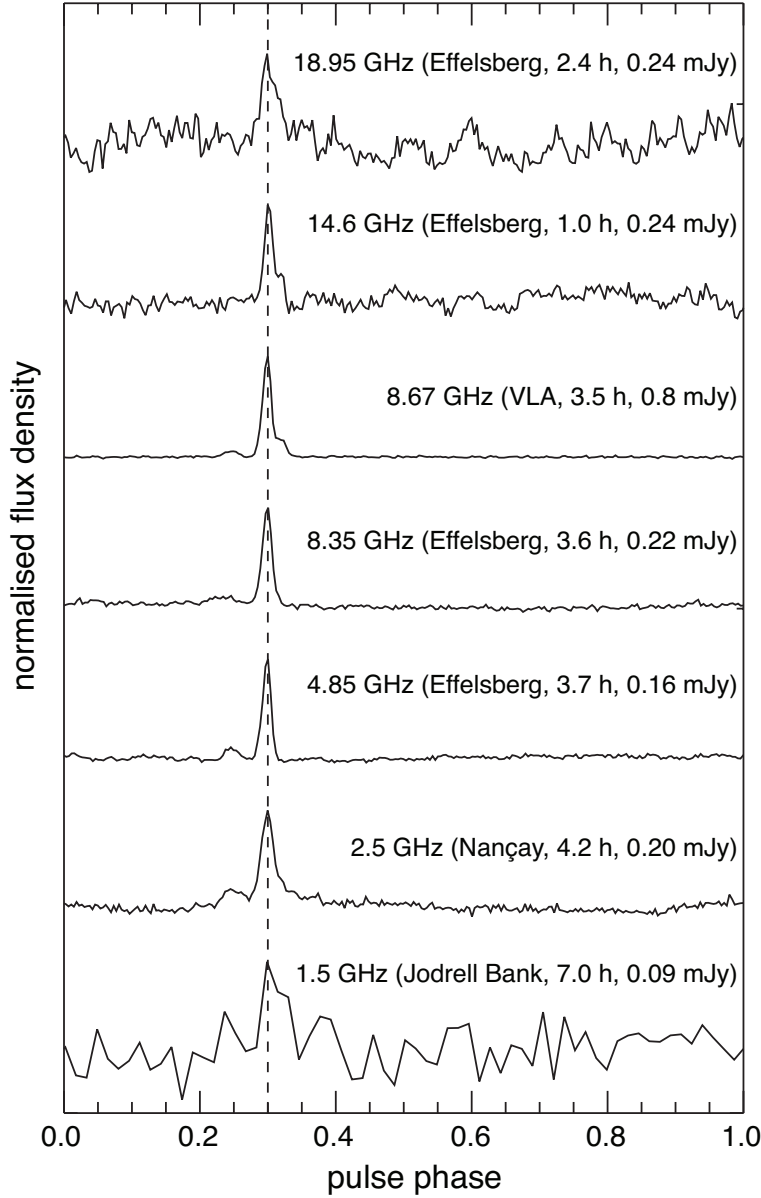}
\caption{\footnotesize Average pulse profiles of PSR J1745-2900 at each of the radio
  frequencies where detections have been made.  All observations have
  been centered on the X-ray position measured with {\em
    Chandra}\cite{rea_atel}. The telescope used, the total observation
  time required to generate the profile, and the average flux density
  is indicated in brackets after the frequency label. In each case the
  profiles have been down-sampled from the original sampling interval
  to 256 phase bins (64 for the Jodrell data), and the peak flux
  density normalised to unity. The profiles have been aligned by the
  peak of the main pulse detected. By measuring accurate arrival times
  of the pulses, we have constructed a coherent timing solution; a
  model which tracks every single rotation of the pulsar. Over the
  period MJD 56414--56426, this model has given measurements of the
  spin period $P=3.76354676(2)$ s and period derivative (spin-down)
  $\dot P = 6.82(3)\times10^{-12}$; uncertainties on the last digit,
  given in brackets, are derived from the one-sigma error of timing
  model fit. Absolute timing from 1.5 to 8.35 GHz, has established
  that the main pulse in each profile is indeed aligned at each
  frequency.}
\end{figure}

\begin{figure}
\includegraphics[scale=0.72]{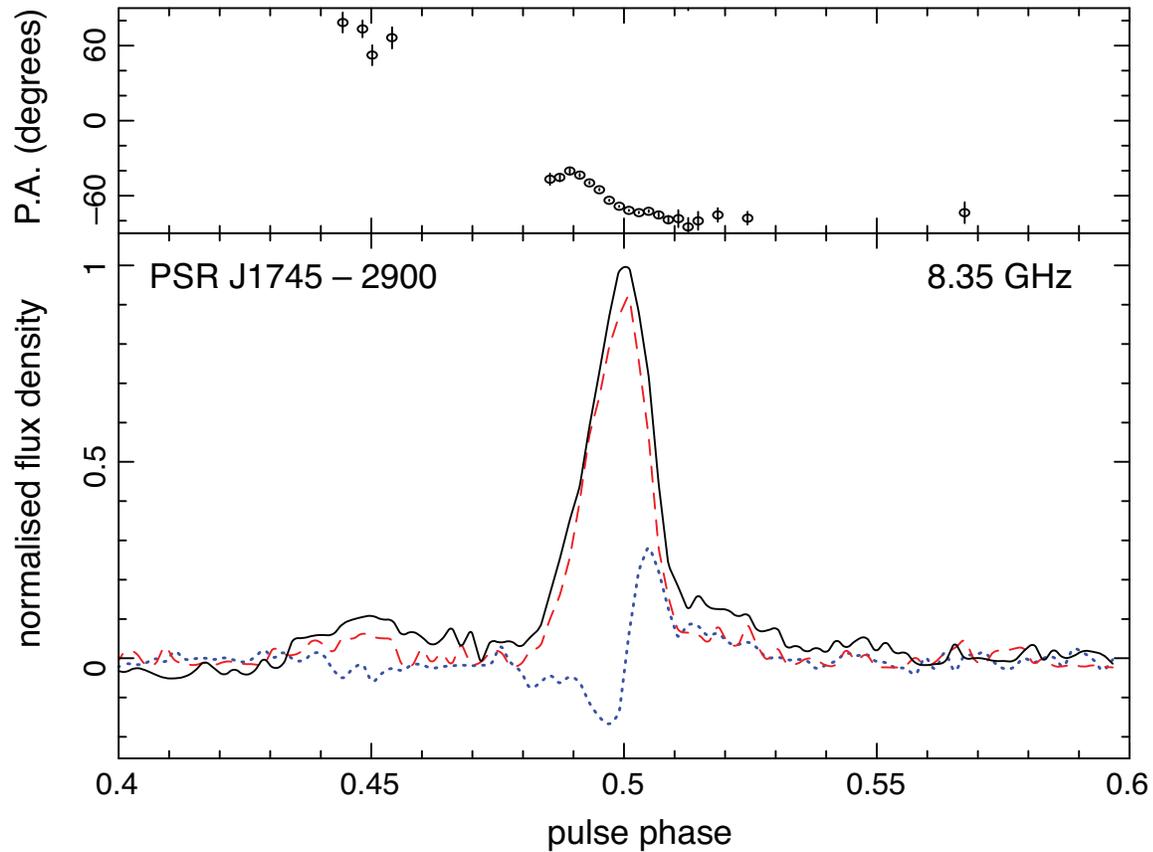}
\caption{\footnotesize Pulse profile of PSR J1745-2900 at 8.35 GHz. After correcting
  for the Faraday rotation of $(-6.696 \pm 0.005) \times 10^{4} \;
  {\rm rad \; m^{-2}}$, we can measure the intrinsic polarization
  across the pulse profile, together with the polarization position
  angle (PA). The degree of linear polarization (red dashed line)
  is nearly 100\%, and a significant amount ($\sim 15\%$) of circular
  polarization (blue dotted line) is also detected. A consistent
  `S'-shape PA swing is measured at each frequency.}
\end{figure}

\begin{figure}
\includegraphics[scale=0.71]{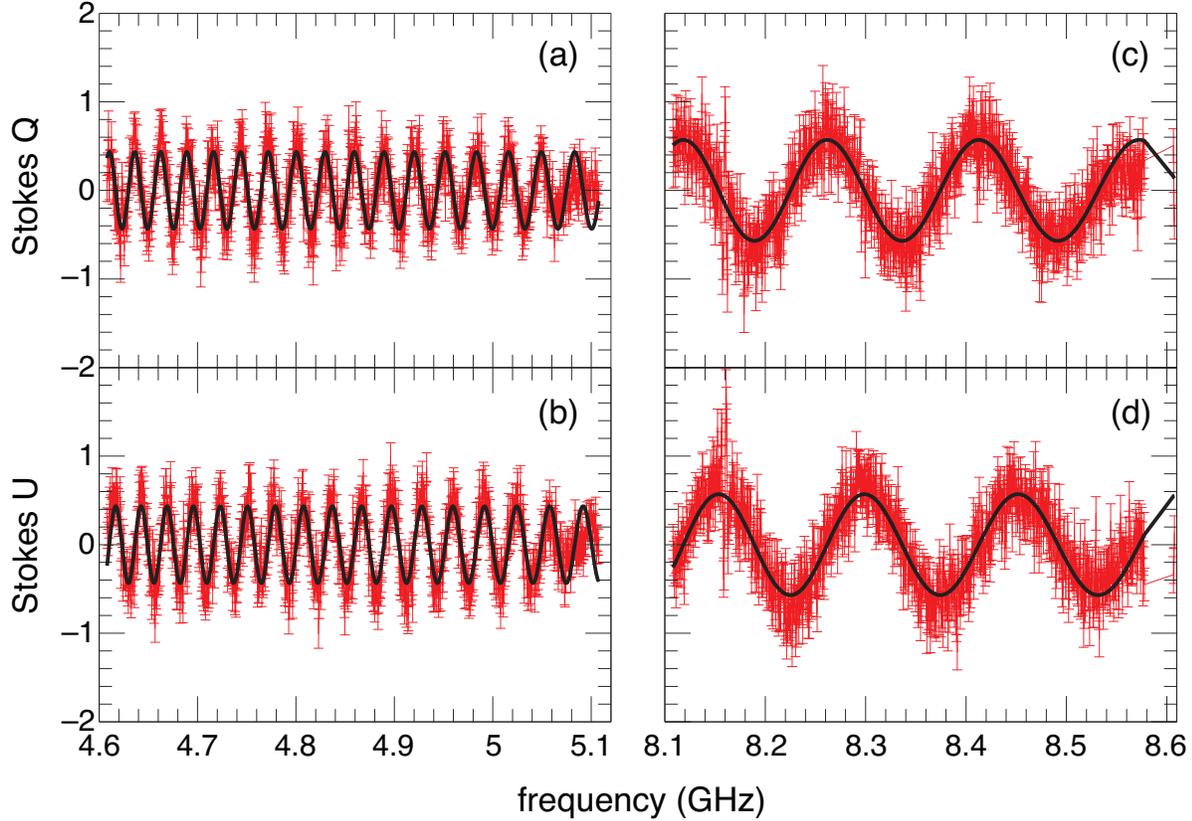}
\caption{\footnotesize Rotation Measure (RM) synthesis analysis for the radio
  polarization of PSR J1745$-$2900. The red points, with one-sigma
  error bars given by the off-pulse baseline r.m.s of the polarization
  profile, present the observed polarized flux density in the Stokes
  parameters Q and U. Note polarization measurements were not possible
  at all frequencies due to hardware limitations. The RM is measured
  by a two-step method. Firstly, we perform the Fourier transformation
  of the polarization intensity to get the RM Faraday spectrum, of
  which the peak is used to find a rough estimation of the RM. Using
  such initial values, we then perform a least-squares fit to the Q
  and U curves to get the RM and its error. The black curves in this
  figure are the model values based on the best fit RM. The sinusoidal
  variation of Q and U due to Faraday rotation is clearly seen across
  the frequency bands centred at 4.85 (panels a. and b.) and 8.35 GHz
  (panels c. and d.). At 2.5 GHz the variation is so severe this
  signature is better seen in the RM spectrum not shown here. RM
  values derived at each frequency band are independently consistent:
  At 2.5 GHz ${\rm RM}=(-6.70 \pm 0.01) \times 10^{4} \; {\rm rad \;
    m^{-2}}$, at 4.85 GHz ${\rm RM}=(-6.694 \pm 0.006) \times 10^{4}\;
  {\rm rad \; m^{-2}}$ and at 8.35 GHz ${\rm RM}=(-6.68 \pm
  0.04)\times 10^{4} \; {\rm rad \; m^{-2}}$. The RM has also been
  measured with the VLA at 8.67 GHz giving $(-6.70 \pm 0.04) \times
  10^{4} \; {\rm rad \; m^{-2}}$. The combined and appropriately
  weighted average is $(-6.696 \pm 0.005) \times 10^{4} \; {\rm rad \;
    m^{-2}}$.}
\end{figure}


\end{document}